\documentclass[12pt]{article}

\newcommand{\mdet}{\mbox{det}}

\newcommand{\barA}{\bar{A}}
\newcommand{\barI}{\bar{I}}
\newcommand{\barR}{\bar{R}}
\newcommand{\barU}{\bar{U}}

\newcommand{\barW}{\bar{W}}
\newcommand{\barX}{\bar{X}}
\newcommand{\barw}{\bar{w}}

\newcommand{\hatA}{\hat{A}}
\newcommand{\hatI}{\hat{I}}
\newcommand{\hatR}{\hat{R}}
\newcommand{\hatU}{\hat{U}}
\newcommand{\hatV}{\hat{V}}
\newcommand{\hatW}{\hat{W}}
\newcommand{\hatX}{\hat{X}}
\newcommand{\hatw}{\hat{w}}

\newcommand{\mbbR}{\mathbb{R}}

\newcommand{\mcC}{\mathcal{C}}

\newcommand{\eqdist}{\stackrel{d}{=}}

\usepackage[latin1]{inputenc}
\usepackage{amsfonts,epsfig,graphicx}
\usepackage{amsmath,amssymb,amsthm}
\usepackage{epsf}
\usepackage{graphics}
\usepackage{amsfonts}
\usepackage{amsmath}
\usepackage{amssymb}
\usepackage[usenames,dvipsnames]{color}
\usepackage{graphicx}
\usepackage{times}

\usepackage{latexsym}

\usepackage{mathbbol}
\usepackage{bbm}

\usepackage[colorlinks,citecolor=blue]{hyperref}
\usepackage{tikz}
\usetikzlibrary{calc,positioning}
\usetikzlibrary{patterns}
\usepackage{bbm}
\usepackage{float}
\usepackage{enumitem}
\usepackage{comment}

\usepackage{MnSymbol}

\usepackage[american voltages, american currents]{circuitikz}

\begin{document}

\begin{center}
{\Large \bf An information-theoretic proof of the Shannon-Hagelbarger theorem}\\
~\\
Venkat Anantharam\\
{\sl EECS Department}\\
{\sl University of California}\\
{\sl Berkeley, CA 94720, U.S.A.}\\
~\\
~\\
{\bf Abstract}
\end{center}

The Shannon-Hagelbarger theorem states that the effective resistance
across any pair of nodes in a resistive network is a concave function of the edge resistances.
We give an information-theoretic proof of this result, building on the theory of the Gaussian
free field. This also allows us to prove an extension of the result to 
determinants of matrices of cross effective resistances.

\section{Introduction}

Consider a connected undirected graph $G := (V,E)$ with no self-loops. 
Here $V$ is a finite set, whose elements will be called vertices or nodes, and
$E$ is a finite set, whose elements will be called edges.
Each edge $e \in E$ is corresponds to an unordered pair $(i,j)$ with
$i, j \in V$, $i \neq j$ and an index in $\{1, \ldots, n_{(i,j)}\}$ for some $n_{(i,j)} \ge 1$
depending only on the unordered pair $(i,j)$ and defined only if there is an edge corresponding
to $(i,j)$.
Thus an edge is between a pair of nodes, but we allow for multiple edges between any pair of nodes.
To each edge $e \in E$ we associate a resistance $R_e > 0$. The resulting weighted graph 
will be called a resistive network.

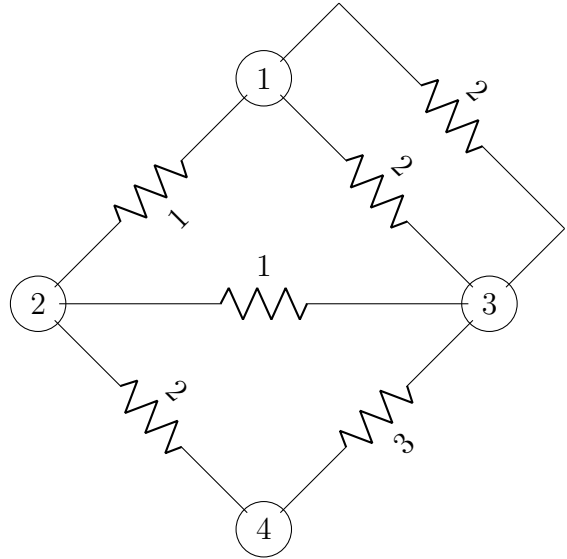
\begin{figure}		\label{fig:network}
\begin{center}

\begin{circuitikz}



   \draw (3,3) node[shape=circle, draw] {$1$};
   \draw (0,0) node[shape=circle, draw] {$2$};
   \draw (6,0) node[shape=circle, draw] {$3$};
   \draw (3,-3) node[shape=circle, draw] {$4$};
   \draw [R = $1$] (2.78,2.78) to (0.22,0.22);
   \draw [R = $2$] (0.22,-0.22) to (2.78,-2.78);
      \draw [R = $2$] (3.22,2.78) to (5.78,0.22);
       \draw [R = $3$] (5.78,-0.22) to (3.22,-2.78);
       \draw [R = $1$] (0.28,0) to (5.72,0); 
       \draw (3.22,3.22) -- (4,4);
       \draw (6.22,0.22) -- (7,1);
       \draw [R = $2$] (4,4) to (7,1);




\end{circuitikz}
 \caption{A resistive network with vertex set $V = \{1, 2, 3, 4\}$ and 
 six edges, the resistance of each of which is indicated. The effective resistances can be computed to be $R^{\rm eff}_{12} = \frac{11}{17}$,
 $R^{\rm eff}_{13} = \frac{11}{17}$, $R^{\rm eff}_{14} = \frac{29}{17}$,
 $R^{\rm eff}_{23} = \frac{10}{17}$, $R^{\rm eff}_{24} = \frac{22}{17}$,
 and $R^{\rm eff}_{34} = \frac{24}{17}$.
 }
\end{center}
\end{figure}

Figure \ref{fig:network} illustrates a resistive network with 
vertex set $V = \{1, 2, 3, 4\}$ and six edges, the resistance of each
of which is indicated.

The Shannon-Hagelbarger theorem \cite{ShanHagel} states that in a resistive network
the effective resistance
across any pair of nodes is a concave function of the edge resistances.
To understand this statement we need to 
recall the concept of effective resistance across a pair of nodes. For this
we need some familiarity with basic
circuit theory notions, which are well-explained, for instance, in the book of Doyle and Snell
\cite{DoyleSnell}. Consider distinct nodes $a, b \in V$ and
imagine that a unit current source is placed across them, injecting current $1$ at node $a$ and
extracting it at node $b$. Kirchhoff's current law (which states that the net current entering each 
node will be $0$) and Kirchhoff's voltage law (which states that the sum of voltage drops around
any circuit will be $0$) will then result in a unique assignment of voltages to each node if we 
set the voltage at node $b$ to $0$. The effective resistance across the pair of distinct nodes $a, b \in V$
is then defined to be the voltage at node $a$. This is just a mathematical statement, of course, the physically
relevant statement being that the effective resistance across the pair distinct nodes $a, b \in V$
is this voltage divided by the unit source current.
From this description, it is straightforward to argue that the same effective resistance results if the
roles of $a$ and $b$ are interchanged, so the effective resistance only depends on the unordered pair $(a,b)$.

In the example of Figure \ref{fig:network} there are $6 = {4 \choose 2}$ effective
resistances to be computed, with their values being as indicated in the caption of the figure.

\section{Existing proofs}

The proof of the Shannon-Hagelbarger theorem in \cite{ShanHagel} 
is circuit-theoretic in nature. It is proposed that a specific
ideal transformer structure be appended to the network
with edge resistances
$(\hatR_e := R_e + \barR_e, e \in E)$, with a switch that, when open,
results in the effective resistance across $(a, b)$ being that
of the network with edge resistances
$(\hatR_e, e \in E)$, denoted by $\hatR^{\rm eff}_{ab}$,
while, when closed, leads to this effective resistance being replaced by the
sum of the effective resistance across $(a,b)$ when the edge resistances are $(R_e, e \in E)$ to that
when the edge resistances are $(\barR_e, e \in E)$, i.e. by $R^{\rm eff}_{ab} + \barR^{\rm eff}_{ab}$.
Here $R_e > 0$ and $\barR_e > 0$ for all $e \in E$. 
It is physically obvious 
that closing the switch results in a smaller effective resistance, 
since it can be thought of, informally, as being equivalent to replacing an infinite
resistance across some edge with zero resistance, and the effective resistance is a monotone increasing
function of the individual edge resistances.

Now, since
the effective resistance scales linearly when all the edge resistances are scaled by the same amount, 
the inequality $\hatR^{\rm eff}_{ab} \ge R^{\rm eff}_{ab} + \barR^{\rm eff}_{ab}$
that was just proved is enough to establish the claimed concavity.
Rather than appealing just to this physical intuition, a circuit-theoretic argument, based on replacing a
two-terminal circuit by its so-called T-structure is given in \cite{ShanHagel} to give a mathematical 
proof of the claimed result. For the details, see \cite{ShanHagel}.

An elementary proof of the Shannon-Hagelbarger theorem
was subsequently 
given by Melvin \cite{Melvin} based on Kirchhoff's current law and the minimum energy principle
of Thomson, 
which, in the scenario where a unit current source is place across the pair of distinct nodes $a, b \in V$ 
in a resistive network, states that among all the resulting current flows that satisfy Kirchhoff's current law
the unique one that satisfies Kirchhoff's voltage law will be the one for 
which the total power dissipated in the network, i.e. $\sum_{e \in E } I_e^2 R_e$, is 
minimized. See \cite[Sec. 3.5]{DoyleSnell} for more details. Here, by a current flow we mean $(I_e, e \in E)$ where 
$I_e \in \mbbR$ is the current along edge $e$ in some fixed orientation (for instance,
fixing some ordering on the vertex set $V$ allows us to think of each edge as having a natural orientation).
Note that since only the square of the current appears in the use of Thomson's principle, the choice of
orientation of the edges is irrelevant. Also note that the resulting effective resistance would 
then just be $R^{\rm eff}_{ab} = \sum_{e \in E } I_e^2 R_e$ for the current flow 
$(I_e, e \in E)$ that minimizes power dissipation, 
because the total power dissipated in the network, i.e. $\sum_{e \in E } I_e^2 R_e$, 
is that dissipated by a unit current
flowing into the effective resistance $R^{\rm eff}_{ab}$.

We may now give a proof of the Shannon-Hagelbarger theorem 
as follows (this is basically the argument in \cite{Melvin}). Given the choices of edge resistances 
$(R_e, e \in E)$ and $(\barR_e, e \in E)$, where $R_e > 0$ and $\barR_e > 0$ for each
$e \in E$, let $\hatR_e := R_e + \barR_e$ for $e \in E$ and also consider the choice
of edge resistances $(\hatR_e, e \in E)$. Let $(I_e, e \in E)$, $(\barI_e, e \in E)$,
and $(\hatI_e, e \in E)$ be the current flows in the respective scenarios that obey Kirchhoff's 
current law and the Thomson minimum energy principle, when a unit current source injects current at 
node $a$ and extracts it at node $b$. Then we have
\begin{eqnarray*}
\hatR^{\rm eff}_{ab} &=& \sum_{e \in E } \hatI_e^2 \hatR_e\\
&=& \sum_{e \in E } \hatI_e^2 R_e + \sum_{e \in E } \hatI_e^2 \barR_e\\
&\ge& \sum_{e \in E } I_e^2 R_e + \sum_{e \in E } \barI_e^2 \barR_e\\
&=& R^{\rm eff}_{ab} + \barR^{\rm eff}_{ab},
\end{eqnarray*}
where the inequality follows from the minimum energy principle. 
After this the theorem follows by the same linear scaling argument used in \cite{ShanHagel}.

\section{A probabilistic point of view}

The proof of the Shannon-Hagelberger theorem in \cite{Melvin}, which is essentially reproduced 
entirely above, is as simple as it gets, so it might seem to be the last word on the story.
Nevertheless, given that Shannon is widely recognized as the creator of information theory, 
one is led to observe that there has not been even a whiff of information theory 
in the discussion so far. The main contribution of this paper is to point out that, from the
appropriate viewpoint, the Shannon-Hagelberger theorem can be thought of as a theorem of
information theory. 

To start with, to every resistive network we can associate a multivariate Gaussian 
random variable $(X_e, e \in E)$, where the components are independent mean-zero Gaussian
random variables with respective variances $R_e$. 

Since we allow for multiple edges, we need to define the notion of a walk in the graph $G$ more carefully
than in the situation without multiple edges. By a walk we mean a sequence 
$w := (v_0, e_0, v_1, e_1, \ldots, v_{n(w)-1}, e_{n(w)-1}, v_{n(w)})$, where $n(w) \ge 1$, $v_k \in V$ for $0 \le k \le n(w)$,
$e_k \in E$ for $0 \le k \le n(w)-1$, and the vertex pair corresponding to $e_k$ is
$(v_k, v_{k+1})$ for $0 \le k \le n(w)-1$. 
A simple closed walk, also called a circuit,
is a walk $c$ with $n(c) \ge 2$ and $v_{n(c)} = v_0$, and where no vertex appears more than once. 
It can be checked that in a circuit no edge can appear more than once.
Let $\mcC$ denote the set of all circuits in the graph. This is a finite set. 

We next pick an arbitrary ordering on $V$, the set of vertices. 
For a walk
$w := (v_0, e_0, v_1, e_1, \ldots, v_{n(w)-1}, e_{n(w)-1}, v_{n(w)})$ 
and $0 \le k \le n(w) - 1$, we 
write $\epsilon_{w,k} = 1$ if $v_{k+1} > v_k$ and 
$\epsilon_{w,k} = -1 $ if $v_k > v_{k+1}$ in the chosen ordering on $V$. 
For any circuit 
$c = (v_0, e_0, v_1, e_1, \ldots, v_{n(c)-1}, e_{n(c)-1}, v_{n(c)} = v_0) \in \mcC$,
let $A_c$ denote the event that 
$\sum_{k=0}^{n(c) -1} \epsilon_{c,k} X_{e_k} = 0$.
We now condition on the occurrence of the event $\cap_{c \in \mcC} A_c$. 
Let $(U_e, e \in E)$ have the law of the conditional distribution of $(X_e, e \in E)$ under this conditioning, 
i.e. 
\[
(U_e, e \in E) \eqdist (X_e, e \in E | \cap_{c \in \mcC} A_c).
\]
Note that $(U_e, e \in E)$ is also a multivariate Gaussian, and that we have 
$\sum_{k=0}^{n(c) -1} \epsilon_{c,k} U_{e_k} = 0$
for all $c = (v_0, e_0, v_1, e_1, \ldots, v_{n(c)-1}, e_{n(c)-1}, v_{n(c)} = v_0) \in \mcC$. 

Let us now pick some $v_* \in V$ and set $\eta_{v_*} = 0$. 
A consequence of the fact that $\sum_{k=0}^{n(c) -1} \epsilon_{c,k} U_{e_k} = 0$
for all $c = (v_0, e_0, v_1, e_1, \ldots, v_{n(c)-1}, e_{n(c)-1}, v_{n(c)} = v_0) \in \mcC$
is that we can define a multivariate Gaussian
$(\eta_v, v \in V)$ by defining
\[
\eta_v := \sum_{k=0}^{n(w) -1} \epsilon_{w,k} U_{e_k}
\]
for any walk $w := (v_0, e_0, v_1, e_1, \ldots, v_{n(w)-1}, e_{n(w)-1}, v_{n(w)})$
such that $v_0 = v_*$ and $v_{n(w)} = v$. The point is that the same $\eta_v$ results, irrespective
of the choice of the walk. 
The multivariate Gaussian $(\eta_v, v \in V)$ is called the Gaussian free field 
associated to the given resistive network. 
As defined, this concept seems to be depend on the choice of $v_*$, but it should be apparent that
what is really important are the differences $\eta_b - \eta_a$ for distinct $a, b \in V$, which will 
be the same irrespective of the choice of $v_*$.
The Gaussian free field is an object that is widely studied in probability theory
in many contexts, see e.g.  \cite[Chap. 6]{Lawler}, \cite[Sec. 2.8]{LPbook}, \cite[Sec. 9.4]{Janson},
\cite[Sec. 1.5.2]{WernerPowell}.

For any distinct pair of vertices $a, b \in V$, we now define $U_{a \to b} := \eta_b - \eta_a$. 
It can be checked that for an edge $e = (i,j)$ we have $U_{i \to j} = U_e$ if $i < j$ and 
$U_{i \to j} = - U_e$ if $j < i$ in the chosen ordering on $V$. Note also that 
$U_{b \to a} = - U_{a \to b}$. It is known that for any distinct pair of vertices $a, b \in V$
the random variable
$U_{a \to b}$ is a mean zero Gaussian random variable with variance $R^{\rm eff}_{ab}$. 
See \cite[Prop. 2.24]{LPbook} for more details.

In the example of Figure \ref{fig:network}, suppose we choose the natural ordering on the vertices and number the $(1,2)$ edge as $1$, the
two $(1,3)$ edges as $2$ and $3$ respectively, the $(2,3)$ edge as
$4$, the $(2,4)$ edge as $5$, and the $(3,4)$ edge as $6$. 
The random variables $(X_1, X_2, X_3, X_4, X_5, X_6)$ are then 
independent Gaussians with mean zero and variances $1, 2, 2, 1, 2, 3$ 
respectively. We can compute the conditional covariance matrix of these random variables conditioned on requiring zero sums over all circuits,
i.e. conditioned on $X_1 + X_4 - X_2 = 0$, $X_2 - X_3 = 0$, and
$X_4 + X_6 - X_5 = 0$. It can be verified that the 
conditional variances of each of the random variables $X_1$ through
$X_6$, subject to this conditioning, equal the respective effective resistances as listed in the caption of the figure.

\section{An information-theoretic proof} \label{sec:it}
We write $h(Z)$ for the differential entropy of the real-valued random variable $Z$.
We write $h(Z|A)$ for the conditional differential entropy of $Z$ conditioned on the event $A$. We write $h(Z|W)$ for the
conditional differential entropy of $Z$ given $W$ when $(Z,W)$ 
are jointly defined in $\mbbR \times \mbbR^n$

We are now in a position to give an information-theoretic proof of the Shannon-Hagelbarger theorem.
Given the choices of edge resistances 
$(R_e, e \in E)$ and $(\barR_e, e \in E)$, where $R_e > 0$ and $\barR_e > 0$ for each
$e \in E$, let $\hatR_e := R_e + \barR_e$ for $e \in E$ and also consider the choice
of edge resistances $(\hatR_e, e \in E)$. 
Consider a multivariate Gaussian 
random variable $(X_e, e \in E)$, where the components are independent mean-zero Gaussian
random variables with respective variances $R_e$ and another
multivariate Gaussian 
random variable $(\barX_e, e \in E)$, where the components are independent mean-zero Gaussian
random variables with respective variances $\barR_e$, and assume that 
$(X_e, e \in E)$ and $(\barX_e, e \in E)$ are independent. 
Define $(\hatX_e, e \in E)$ by $\hatX_e = X_e + \barX_e$ for $e \in E$. 
Then in $(\hatX_e, e \in E)$ the components are independent mean-zero Gaussian
random variables with respective variances $\hatR_e$.

For any circuit 

$c = (v_0, e_0, v_1, e_1, \ldots, v_{n(c)-1}, e_{n(c)-1}, v_{n(c)} = v_0) \in \mcC$,
let $A_c$ denote the event that 
$\sum_{k=0}^{n(c) -1} \epsilon_{c,k} X_{e_k} = 0$, 
let $\barA_c$ denote the event that 
$\sum_{k=0}^{n(c) -1} \epsilon_{c,k} \barX_{e_k} = 0$, and
let $\hatA_c$ denote the event that 
$\sum_{k=0}^{n(c) -1} \epsilon_{c,k} \hatX_{e_k} = 0$.

We may then define $(U_e, e \in E) \eqdist (X_e, e \in E | \cap_{c \in \mcC} A_c)$,
$(\barU_e, e \in E) \eqdist (\barX_e, e \in E | \cap_{c \in \mcC} \barA_c)$,
and $(\hatU_e, e \in E) \eqdist (\hatX_e, e \in E | \cap_{c \in \mcC} \hatA_c)$,
with $(U_e, e \in E)$ being independent of $(\barU_e, e \in E)$.

Note now that for any distinct vertices $a, b \in V$ we have,
for any walk $w := (v_0, e_0, v_1, e_1, \ldots, v_{n(w)-1}, e_{n(w)-1}, v_{n(w)})$
with $v_0 = a$ and $v_{n(w)} = b$, the following:
\begin{eqnarray*}
h(\hatU_{a \to b}) &=& h( \sum_{k=0}^{n(w) -1} \epsilon_{w,k} \hatU_{e_k})\\
&=& h( \sum_{k=0}^{n(w) -1} \epsilon_{w,k} \hatX_{e_k} | \cap_{c \in \mcC} \hatA_c )\\
&=& h( \sum_{k=0}^{n(w) -1} \epsilon_{w,k} (X_{e_k} + \barX_{e_k})  | \cap_{c \in \mcC} \hatA_c )\\
&\ge& h( \sum_{k=0}^{n(w) -1} \epsilon_{w,k} (X_{e_k} + \barX_{e_k})  | \cap_{c \in \mcC} A_c, \cap_{c \in \mcC} \barA_c )\\
&\stackrel{(a)}{=}& h( \sum_{k=0}^{n(w) -1} \epsilon_{w,k} U_{e_k} + \sum_{k=0}^{n(w) -1} \epsilon_{w,k} \barU_{e_k})\\
&=& h( U_{a \to b} + \barU_{a \to b}),
\end{eqnarray*}
where the inequality follows because the conditioning reduces entropy, and step (a) follows from the 
asssumption that $(X_e, e \in E) \amalg (\barX_e, e \in E)$, the $A_c$ are defined purely in terms of 
$(X_e, e \in E)$ and the $\barA_c$ purely in terms of $(\barX_e, e \in E)$, and from the assumption that
$(U_e, e \in E) \amalg (\barU_e, e \in E)$. Since we are 
conditioning on an event, we need an argument for why the
conditioning reduces entropy; this is given in the appendix.

Now, $U_{a \to b}$ and $\barU_{a \to b}$ are independent mean-zero Gaussian random variables with 
variances $R^{\rm eff}_{ab}$ and $\barR^{\rm eff}_{ab}$ respectively, while 
$\hatU_{a \to b}$ is a mean-zero Gaussian random variable with variance $\hatR^{\rm eff}_{ab}$.
It follows that $\hatR^{\rm eff}_{ab} \ge R^{\rm eff}_{ab} + \barR^{\rm eff}_{ab}$,
after which an argument based on the linear scaling of effective resistances when all the edge resistances
are scaled by the same amount completes the information-theoretic proof of the Shannon-Hagelbarger theorem.

\section{Concluding remarks}

The purpose of writing this article was mostly to go on a whimsical stroll.
What we have discovered is the curious fact that a theorem co-authored
by Shannon, which on the face of it has no connections with information theory, is really a 
theorem in information theory after all.

However, the new proof scheme also provides additional information,
generalizing the Shannon-Hagelbarger theorem, which, to the best of our knowledge, appears to be new. For instance, given four nodes $a, b, c, d \in V$, not necessarily distinct, we may define a cross effective resistance 
between $a \to b$ and $c \to d$ by 
$R^{\rm eff}_{a \to b, c \to d} := E[U_{a \to b} U_{c \to d}]$.
In the resistive network this has the interpretation of the voltage drop
from node $c$ to node $d$ when a unit current is injected at node $a$ and 
extracted at node $b$, see \cite[Prop. 2.24 (ii)]{LPbook}.

With the setting and notation as in Section \ref{sec:it}, a similar calculation would
then lead to the inequality
\[
h( \hatU_{a \to b}, \hatU_{c \to d}) 
\ge h(U_{a \to b} + \barU_{a \to b}, U_{c \to d}, \barU_{c \to d}),
\]
for joint entropies. This then corresponds to the inequality
\begin{eqnarray}	 \label{eq:two}
&& \mdet \begin{bmatrix} 
\hatR^{\rm eff}_{ab} & \hatR^{\rm eff}_{a \to b, c \to d}\\
\hatR^{\rm eff}_{a \to b, c \to d} & \hatR^{\rm eff}_{cd}
\end{bmatrix} 
\ge \\
&&~~~
\mdet \begin{bmatrix} 
R^{\rm eff}_{ab} + \barR^{\rm eff}_{ab} & R^{\rm eff}_{a \to b, c \to d}
+ \barR^{\rm eff}_{a \to b, c \to d}\\
 R^{\rm eff}_{a \to b, c \to d}
+ \barR^{\rm eff}_{a \to b, c \to d} & R^{\rm eff}_{cd} +
\barR^{\rm eff}_{cd}
\end{bmatrix}. \nonumber
\end{eqnarray}

Note that $\begin{bmatrix} 
R^{\rm eff}_{ab} & R^{\rm eff}_{a \to b, c \to d}\\
R^{\rm eff}_{a \to b, c \to d} & R^{\rm eff}_{cd}
\end{bmatrix}$ is positive semidefinite,
because it is the covariance matrix of the random vector
$\begin{bmatrix} U_{a \to b}\\ U_{c \to d} \end{bmatrix}$.
Note that the square root of 
$\mdet \begin{bmatrix} 
R^{\rm eff}_{ab} & R^{\rm eff}_{a \to b, c \to d}\\
R^{\rm eff}_{a \to b, c \to d} & R^{\rm eff}_{cd}
\end{bmatrix}$
scales linearly when all the resistances in the network are scaled by
the same constant. 
Recall Minkowski's
determinantal inequality for positive semidefinite matrices, 
which states that for $n \times n$ positive semidefinite matrices
$K_1$ and $K_2$ we have 
$(\mdet( K_1 + K_2) )^\frac{1}{n} 
\ge (\mdet K_1)^\frac{1}{n} +  (\mdet K_2)^\frac{1}{n}$.
Putting these together, we can conclude from the inequality 
\eqref{eq:two} that 
$\sqrt{ \mdet \begin{bmatrix} 
R^{\rm eff}_{ab} & R^{\rm eff}_{a \to b, c \to d}\\
R^{\rm eff}_{a \to b, c \to d} & R^{\rm eff}_{cd}
\end{bmatrix}}$ is
a concave function of the individual edge resistances. 
For a proof of the Minkowski determinantal inequality based on the
entropy power inequality, see \cite[Thm. 7]{CTDet}.

A similar argument for multiple choices of ordered pairs
of vertices shows that the $n$-th root of the determinant of 
cross effective resistances corresponding to any $n$ such 
ordered pairs is a concave function of all the individual edge resistances
in the resistive network.


\section*{Acknowledgment}

This research was supported by the NSF grants
CCF-1901004 and CIF-2007965.
Thanks to Devon Ding for reading a draft of this document for a 
sanity check.

\section*{Appendix}

We prove the claim that the conditioning reduces entropy in the 
calculation in the main text.
Suppose $W$ and $\barW$
are mean zero Gaussian random variables in 
$\mbbR^n$ with
$W \amalg \barW$, and let $\hatW := W + \barW$. 
Let $\hatV$ be any scalar random variable such that
$(\hatV, W, \barW)$ is jointly Gaussian.
Then we have the inequality
$h(\hatV | \hatW = 0) \ge h( \hatV | W = 0, \barW = 0)$
for conditional differential entropies. To see this,
note that $h(V| \hatW = \hatw)$ is the same for all
$\hatw$, and $h( \hatV | W = w, \barW = \barw)$ 
is the same for all $(w, \barw)$.
Thus $h(\hatV | \hatW = 0) \ge h( \hatV | W = 0, \barW = 0)$ follows from
$h(\hatV | \hatW) \ge h(\hatV | W, \barW)$, which is what one
usually means when saying that conditioning reduces entropy.

\end{document}